\documentclass[aps,prl,twocolumn,a4paper,showpacs,nofootinbib,showpacs]{revtex4-1}
\usepackage{amssymb}
\usepackage{amsmath}
\usepackage{epsfig}
\usepackage{amsfonts}
\usepackage{color}
\usepackage{float}
\usepackage{latexsym}
\usepackage{mathrsfs}
\usepackage{feynmf}
\usepackage{balance}

\begin{document}
\title{ Measurement-induced nonlocality for an arbitrary bipartite state  }
\author{Sayyed Yahya Mirafzali$^{1}$}
\email{yahya.mirafzali@stu-mail.um.ac.ir}
\author {Iman Sargolzahi$^{2}$}
\email{sargolzahi@gmail.com}
\author{ Ali Ahanj$^{3,4}$}
\email{ahanj@ipm.ir}
\author{Kurosh Javidan$^{1}$}
\email{javidan@um.ac.ir}
\author{Mohsen Sarbishaei$^{1}$}
\email{sarbishei@um.ac.ir}
\affiliation{$^{1}$Department of Physics, Ferdowsi University of Mashhad,  Mashhad, Iran. \\
$^{2}$Department of Physics, University of Neyshabur, Neyshabur, Iran.}
\affiliation{$^{3}$ Department of Physics, Khayyam Higher Education Institute (KHEI),  Mashhad, Iran}
\affiliation{$^{4}$ School of Physics, Institute for Research in Fundamental Science (IPM), P. O. Box 19395-5531, Tehran, Iran.}

\begin{abstract}

Measurement-induced nonlocality is a measure of nonlocalty introduced by Luo and Fu [Phys. Rev. Lett \textbf{106}, 120401 (2011)]. In this paper, we study the problem of evaluation of Measurement-induced nonlocality (MIN) for an arbitrary $m\times n$ dimensional bipartite density matrix $\rho$ for the case where one of its reduced density matrix, $\rho^{a}$, is degenerate (the nondegenerate case was explained in the preceding reference). Suppose that, in general, $\rho^{a}$ has $d$ degenerate subspaces with dimension $m_{i}  (m_{i} \leq m , i=1, 2, ..., d)$. We show that according to the degeneracy of $\rho^{a}$, if we expand $\rho$ in a suitable basis, the evaluation of MIN for an $m\times n$ dimensional state $\rho$, is degraded to finding the MIN in the $m_{i}\times n$ dimensional subspaces of state $\rho$. This method can reduce the calculations in the evaluation of MIN. Moreover, for an arbitrary $m\times n$ state $\rho$ for which $m_{i}\leq 2$, our method leads to the exact value of the MIN. Also, we obtain an upper bound for MIN which can improve the ones introduced in the above mentioned reference. In the final, we explain the evaluation of MIN for $3\times n$ dimensional states in details.

\end{abstract}


\maketitle

\section{Introduction}

Quantum mechanics is a nonlocal theory. The principle of locality states that the properties of one particle can not be affected by another particle that is sufficiently far away. Nonlocality in quantum mechanics, at least, has two different aspects~\cite{1}. On of these aspects arises in the Aharonov-Bohm effect. The Aharonov-Bohm effect is nonlocal in the sense that the electromagnetic field influences an electron in a region where the field is zero. The other aspect of nonlocality in quantum mechanics appears when one performs local measurements on spatially separated systems. These nonlocal effects result from the fact that the local measurements can alter the overall state of a multipartite quantum system. Quantum nonlocality usually refers to this aspect of nonlocality and is studied often in the context of Bell inequalities. Bell inequalities are mathematical relations setting conditions on the results of measurements made on separated systems. These inequalities are satisfied by any local hidden variable theory, but they may be violated by quantum mechanics. This is the very feature of quantum mechanics that is usually mentioned as quantum nonlocality.

Recently, S. Luo and S. Fu in Refs.~\cite{2,3} studied the latter aspect of nonlocality based on an approach different from the violation of Bell inequalities. They have used this idea that, in general, measurement in quantum mechanics causes disturbance. If one performs local measurements that do not disturb the states of the subsystems, then any disturbance in the system's overall state, can be attributed to the genuine nonlocal features of the system. The disturbance caused by local measurements leaving the states of the subsystems invariant can be quantified by distance between the overall pre and post-measurement states of the system. Based on this idea, they defined the measurement-induced nonlocality (MIN) as the maximum distance between the bipartite state $\rho$ and its post-measurement state, where the maximum is taken over all the von Neumann local measurements which do not disturb the local state $\rho^{a}=tr_{b}\rho$~\cite{3}.

The MIN is a manifestation of quantum nonlocality besides the violation of Bell inequalities. It is different from entanglement and discord, although for any pure state it coincide with the geometric measure of quantum discord and the square of concurrence~\cite{2,3}. MIN can provide a novel classification scheme for bipartite states, and it can also be a quantum resource in quantum information processing, although there is no operational interpretation for it.

The analytical formulas of MIN for any state with non-degenerate local state $\rho^{a}=tr_{b}\rho$, arbitrary dimensional bipartite pure states and $2\times n$ dimensional mixed state was obtained in Refs.~\cite{2,3}. In addition, in Ref.~\cite{3}, a tight upper bound on the MIN of an arbitrary $m\times n$ dimensional state is derived. The necessary and sufficient conditions for a state to have nullity of MIN were studied in Ref.~\cite{4}. In this paper, we study the evaluation of MIN for an arbitrary $m\times n$ dimensional bipartite state. We introduce a method that can reduce the calculations in the evaluation of MIN. Using this method, we obtain an upper bound for MIN which can improve the ones introduced in Ref.~\cite{3}.

The paper is organized as follows. In the next section (Sec. II), we review the results of Ref.~\cite{3}. Our main results are given in Sec. III. In Sec. IV, $3\times n$ dimensional states are studied. Finally, we give some conclusions in Sec. V.


\section{measurement-induced nonlocality}

Consider a bipartite state $\rho$ defined in Hilbert space $H^{a}\otimes H^{b}$ where $H^{a}$ ($H^{b}$) is the Hilbert space of part $a$ ($b$) with dimension $m$ ($n$). Suppose this state is shared between two distant parties $A$ and $B$. Assume that party $A$ performs a non-selective local von Neumann measurement on his (her) part, then the state of system, $\rho$, changes to $\Pi^{a}(\rho)=\sum_{k}(\Pi^{a}_{k}\otimes 1^{b})\rho(\Pi^{a}_{k}\otimes 1^{b})$ where $\Pi^{a}=\lbrace\Pi^{a}_{k}\rbrace $ is a set of orthogonal, one dimensional projectors summing to the identity. Then, the measurement-induced nonlocality is defined as~\cite{3}:
\begin{equation}
N(\rho)=\max_{\Pi^{a}}\Vert\rho -\Pi^{a}(\rho)\Vert^{2}\,.
\label{1}
\end{equation}
where the maximum is taken over all the von Neumann measurements $\Pi^{a}=\lbrace\Pi^{a}_{k}\rbrace $ which do not disturb $\rho^{a}$ locally, that is, $\sum_{k}\Pi^{a}_{k}\rho^{a}\Pi^{a}_{k}=\rho^{a}$ and $\Vert.\Vert^{2}$ is the Hilbert Schmidt distance defined as $\Vert X\Vert^{2}=tr X^{\dagger}X$. The maximum in definition of MIN is used to capture all the nonlocal effects that can be induced (indicated) by local measurements.

When $\rho^{a}$ is non-degenerate with spectral decomposition $\rho^{a}=\sum_{k}\lambda_{k}\vert k\rangle\langle k\vert$, then the only von Neumann measurement that does not disturb $\rho^{a}$ is $\lbrace\Pi^{a}_{k}=\vert k\rangle\langle k\vert\rbrace$, so in this case, the maximum in Eq. (\ref{1}) is not necessary. In Refs.~\cite{2,3}, some other basic properties of the MIN, are listed.

Any $m\times n$ bipartite state $\rho$ can be expanded in terms of $\lbrace X_{0}=I/\sqrt{m}, X_{i}: i=1,...,m^{2}-1\rbrace$ and $\lbrace Y_{0}=I/\sqrt{n}, Y_{j}: j=1,...,n^{2}-1\rbrace$ as
\begin{align}
\rho=\dfrac{1}{\sqrt{mn}}\dfrac{I}{\sqrt{m}}\otimes\dfrac{I}{\sqrt{n}}+\sum_{i=1}^{m^{2}-1}x_{i}X_{i}\otimes\dfrac{I}{\sqrt{n}}\qquad\qquad\notag \\
+\dfrac{I}{\sqrt{m}}\otimes\sum_{j=1}^{n^{2}-1}y_{j}Y_{j}+\sum_{i=1}^{m^{2}-1}\sum_{j=1}^{n^{2}-1}t_{ij}X_{i}\otimes Y_{j}.
\label{2}
\end{align}
where $\lbrace X_{i}: i=1,...,m^{2}-1\rbrace$ and $\lbrace Y_{j}: j=1,...,n^{2}-1\rbrace$ are traceless Hermitian operators satisfying the conditions $trX_{i}X_{i'}=\delta_{ii'}$ and $trY_{j}Y_{j'}=\delta_{jj'}$ respectively. It was shown that~\cite{3} MIN can be expressed as:
\begin{align}
 N(\rho)=trTT^{t}-\min_{A} trATT^{t}A^{t}\,.\qquad
\label{3}
\end{align}
where $T=(t_{ij})$ is an $(m^{2}-1)\times(n^{2}-1)$ dimensional matrix and $A=(a_{ki})$ is an $m\times(m^{2}-1)$ dimensional matrix with $a_{ki}=tr(\Pi_{k}^{a}X_{i})$. Here $\lbrace\Pi_{k}^{a}=\vert k\rangle\langle k\vert\rbrace$ is any von Neumann measurement leaving $\rho^{a}$ invariant. In Eq. (\ref{3}) $``t"$ denotes transpose of matrices.

Defining $a_{k0}=tr(\vert k\rangle\langle k\vert X_{0})=1/\sqrt{m}$, then $\lbrace a_{ki}: i=0, 1, ..., m^{2}-1\rbrace$ are the coefficients for expansion of the operator $\vert k\rangle\langle k\vert$ in terms of $\lbrace X_{i}: i=0,...,m^{2}-1\rbrace$ and thus
\begin{align}
\sum_{i=0}^{m^{2}-1} a_{ki}a_{k^{'}i}=tr\vert k\rangle\langle k\vert k^{'}\rangle\langle k^{'}\vert =\delta_{kk^{'}}.
\label{4}
\end{align}
 where $k, k^{'}=1, 2, ..., m$. The author in Ref.~\cite{3}, considering only this latter constraint, obtained the following upper bound for MIN:
\begin{align}
N(\rho)\leq\sum_{i=1}^{m^{2}-m}\lambda_{i}\,.
\label{5}
\end{align}
where the $\lbrace\lambda_{i}: i=1,..., m^{2}-1\rbrace$ are the eigenvalues of the matrix $TT^{t}$, arranged in decreasing order. In addition,  in Ref.~\cite{3}, the analytical formulas of MIN for any dimensional pure state and $2\otimes n$ dimensional mixed state was obtained.

In the next section, we study the problem of evaluation of MIN for an arbitrary $m\times n$ dimensional bipartite density matrix $\rho$ where its reduced density matrix, $\rho^{a}$, has $d$ degenerate subspaces with dimension $m_{i}$.


\section{main results}
In order to evaluate the MIN, at first, we focus on the constraints in the optimization problem in Eq. (\ref{1}). In this equation, the maximum is taken over von Neumann measurements leaving $\rho^{a}$ invariant. A von Neumann measurement is defined via the set  $\Pi^{a}=\lbrace\Pi_{k}^{a}\rbrace$ where the set $\Pi^{a}=\lbrace\Pi_{k}^{a}\rbrace$ is a set of orthogonal, one-dimensional projection operators summing to identity i. e. $\Pi_{k}^{a}\Pi_{k'}^{a}=\delta_{kk'}\Pi_{k}^{a}$, $tr\Pi_{k}^{a}=1$ and $\sum_{k=1}^{m}\Pi_{k}^{a}=I$. These relations set some conditions on $\lbrace a_{ki}\rbrace$. If we expand $\Pi_{k}^{a}$ in terms of $\lbrace X_{i}: i=0, 1, ..., m^{2}-1\rbrace$ i. e. $\Pi_{k}^{a}=\sum_{i=0}^{m^{2}-1} a_{ki}X_{i}$, then from the relation $\Pi_{k}^{a}\Pi_{k'}^{a}=\delta_{kk'}\Pi_{k}^{a}$ we deduce:
\begin{align}
\sum_{i, i'=0}^{m^{2}-1}a_{ki}a_{k'i'}X_{i}X_{i'}=\delta_{kk'}\sum_{i=0}^{m^{2}-1}a_{ki}X_{i}\,.
\label{6}
\end{align}
In addition, from $\sum_{k=1}^{m}\Pi_{k}^{a}=I$ we have $\sum_{i}\sum_{k}a_{ki}X_{i}=I=\sqrt{m}X_{0}$. So from the orthonormality of $\lbrace X_{i}\rbrace$ we obtain
\begin{align}
\sum_{k=1}^{m}a_{ki}=0,  \qquad\quad   i=1, 2, ..., m^{2}-1\,.
\label{7}
\end{align}
Beside the previous constraints, the invariance of $\rho^{a}$ under von Neumann measurements dictates the relation $\rho^{a}=\sum_{k}\Pi_{k}^{a}\rho^{a}\Pi_{k}^{a}$. This relation is fulfilled if and only if $\Pi^{a}=\lbrace\Pi_{k}^{a}\rbrace$ is an spectral projections of $\rho^{a}$, that is $\Pi^{a}=\lbrace\Pi_{k}^{a}=\vert k\rangle\langle k\vert\rbrace$ where $\vert k\rangle$ are eigenvectors of $\rho^{a}$. If $\rho^{a}$ is non-degenerate, then the $\lbrace\vert k\rangle\rbrace$ is unique and so the maximum in Eq. (\ref{1}) is not necessary. On the other hand, when $\rho ^{a}$ is degenerate, any linear combination of eigenvectors corresponding to the same eigenvalue, is also an eigenvector of $\rho^{a}$. But still, for every non-degenerate eigenvalues of $\rho^{a}$ there is a unique eigenvector. So, for an arbitrary $\rho$, taking maximum over non-degenerate subspaces of $\rho^{a}$ in Eq. (\ref{1}) is not necessary and only degenerate ones contribute to the maximum in this equation. If the constraints in Eqs. (\ref{6}), (\ref{7}) can be divided  into constraints  for each degenerate subspace of $\rho^{a}$, independent of the other subspaces of $\rho^{a}$, then the maximum in Eq. (\ref{1}) is degraded to the sum of the maximums for every degenerate subspaces of $\rho^{a}$ separately. This can be done by a suitable construction of $\lbrace X_{i}\rbrace$ from the eigenvectors of $\rho^{a}$ as we do now.

We expand an arbitrary $m\times n$ dimensional state $\rho$ in the form
\begin{align}
\rho=\sum_{i=1}^{m^{2}}\sum_{j=0}^{n^{2}-1}c_{ij}X'_{i}\otimes Y_{j}\,.
\label{8}
\end{align}
where the set $\lbrace Y_{j}\rbrace$ are the same as the one used in Eq. (\ref{2}) and the set $\lbrace X'_{i}\rbrace$ is constructed from the eigenvectors of $\rho^{a}$ as follows. Let us first introduce the notation we use in the next relations. We denote the spectral decomposition of $\rho^{a}$ by $\rho^{a}=\sum_{r=1}^{d}e_{r}\sum_{s=1}^{m_{r}}\vert k_{rs}\rangle\langle k_{rs}\vert+\sum_{r=1}^{d'}e'_{r}\vert k_{r}\rangle\langle k_{r}\vert$ where $\lbrace e_{r}: r=1, ..., d\rbrace$ ($\lbrace e'_{r}: r=1, ..., d'\rbrace$) are the degenerate (non-degenerate) eigenvalues of $\rho^{a}$. In this relation we assume that $\rho^{a}$ has $d+d' (1\leq d+d'\leq m)$ different eigenvalues and each eigenvalue $e_{r}$ is repeated $m_{r}$ times $(2\leq m_{r}\leq m, d'+\sum_{r=1}^{d}m_{r}=m)$. Also, the degenerate eigenvectors of $\rho^{a}$ are denoted by $\vert k_{rs}\rangle$ where the first index specifies the corresponding eigenvalue and the second one is related to the degeneracy of the eigenvalue specified by the index $``r"$. The non-degenerate eigenvectors of $\rho^{a}$ are represented as $\vert k_{r}\rangle$.

At first, we recall that the set $\lbrace\vert k_{rs}\rangle\rbrace$ is not unique and any linear combination of $\vert k_{rs}\rangle$ with the same eigenvalue is also an eigenvector of $\rho^{a}$. For the construction of $\lbrace X'_{i}\rbrace$, we choose one of these sets which we denote as $\lbrace\vert k'_{rs}\rangle\rbrace$. We now construct from $\lbrace\vert k'_{rs}\rangle\langle k'_{rs'}\vert: s,s'=1, ...,m_{r}\rbrace$ the operators $\lbrace X'_{i}: i=B_{r-1}+2, ...,B_{r}\rbrace$, where $B_{r}=m_{1}^{2}+...+m_{r}^{2}$ and $B_{0}=0$, in such a way that they form a set of traceless Hermitian operators satisfying the condition $trX'_{i}X'_{j}=\delta_{ij}$ and we set $X'_{1+B_{r-1}}=\frac{1}{\sqrt{m_{r}}}\sum_{s=1}^{m_{r}}\vert k'_{rs}\rangle\langle k'_{rs}\vert$. Moreover, we put $\lbrace X'_{B_{d}+r}=\vert k_{r}\rangle\langle k_{r}\vert\rbrace$ and the set $\lbrace X'_{i}: i=D, ..., m^{2}\rbrace$ ($B_{d}+d'+1=D$) are constructed from the mixture of $\vert k'_{rs}\rangle\langle k_{r'}\vert$ such that  the relations $ X'_{i}=X_{i}^{'\dagger}$, $trX'_{i}X'_{j}=\delta_{ij}$, $tr(\vert k'_{rs}\rangle\langle k'_{rs}\vert X'_{i})=0$ and $tr(\vert k_{r}\rangle\langle k_{r}\vert X'_{i})=0$ are satisfied.

Now, we evaluate MIN using the expansion of $\rho$ in the new basis. As stated before, the $\Pi^{a}=\lbrace\Pi_{k}^{a}\rbrace$ used in Eq. (\ref{1}) must be an spectral projections of $\rho^{a}$ i. e. in our notation $\Pi^{a}=\lbrace\Pi_{k_{rs}}^{a}=\vert  k_{rs}\rangle\langle k_{rs}\vert\rbrace +\lbrace\Pi_{k_{r}}^{a}=\vert  k_{r}\rangle\langle k_{r}\vert\rbrace$~\cite{5}. After performing von Neumann measurement $\Pi^{a}$ leaving $\rho^{a}$ invariant, the state $\rho$ changes to $\Pi^{a}(\rho)=\sum_{r=1}^{d}\sum_{s=1}^{m_{r}}(\Pi_{k_{rs}}^{a}\otimes 1)\rho(\Pi_{k_{rs}}^{a}\otimes 1)+\sum_{r=1}^{d'}(\Pi_{k_{r}}^{a}\otimes 1)\rho(\Pi_{k_{r}}^{a}\otimes 1)$. Using the Eq. (\ref{8}) we can write $\Pi^{a}(\rho)=\sum_{r=1}^{d}\sum_{s=1}^{m_{r}}\sum_{i=1}^{m^{2}}\sum_{j=0}^{n^{2}-1}c_{ij}\Pi_{k_{rs}}^{a}X'_{i}\Pi_{k_{rs}}^{a}\otimes Y_{j}+\sum_{r=1}^{d'}\sum_{i=1}^{m^{2}}\sum_{j=0}^{n^{2}-1}c_{ij}\Pi_{k_{r}}^{a}X'_{i}\Pi_{k_{r}}^{a}\otimes Y_{j}$. From the construction of $\lbrace X'_{i}\rbrace$ we have $\lbrace\Pi_{k_{rs}}^{a}X'_{i}\Pi_{k_{rs}}^{a}=0: i\neq F_{r}, ..., B_{r}\rbrace$ and $\lbrace\Pi_{k_{r}}^{a}X'_{i}\Pi_{k_{r}}^{a}=0: i\neq B_{d}+1, ..., B_{d}+d'\rbrace$ where $F_{r}=B_{r-1}+1$. Thus, we conclude that $\Pi^{a}(\rho)=\sum_{r=1}^{d}\sum_{s=1}^{m_{r}}\sum_{i=F_{r}}^{B_{r}}\sum_{j=0}^{n^{2}-1}c_{ij}\Pi_{k_{rs}}^{a}X'_{i}\Pi_{k_{rs}}^{a}\otimes Y_{j}+\sum_{i=B_{d}+1}^{B_{d}+d'}\sum_{j=0}^{n^{2}-1}c_{ij}X'_{i}\otimes Y_{j}$. Therefore $\rho-\Pi^{a}(\rho)=\sum_{r=1}^{d}\sum_{i=F_{r}}^{B_{r}}\sum_{j=0}^{n^{2}-1}c_{ij}(X'_{i}-\sum_{s=1}^{m_{r}}\Pi_{k_{rs}}^{a}X'_{i}\Pi_{k_{rs}}^{a})\otimes Y_{j}+\sum_{i=D}^{m^{2}}\sum_{j=0}^{n^{2}-1}c_{ij}X'_{i}\otimes Y_{j}$. On the other hand, from Eq. (\ref{8}) and the invariance of $\rho^{a}$ under the von Neumann measurement we get $\rho^{a}=\sum_{i=1}^{m^{2}}c_{i0}X'_{i}\sqrt{n}$ and  $\rho^{a}=\sum_{r=1}^{d}\sum_{s=1}^{m_{r}}\Pi_{k_{rs}}^{a}\rho^{a}\Pi_{k_{rs}}^{a}+\sum_{r=1}^{d'}\Pi_{k_{r}}^{a}\rho^{a}\Pi_{k_{r}}^{a}
$ respectively. Consequently, we have $\sum_{i=1}^{m^{2}}c_{i0}X'_{i}=\sum_{r=1}^{d}
\sum_{s=1}^{m_{r}}\sum_{i=F_{r}}^{B_{r}}c_{i0}\Pi_{k_{rs}}^{a}X'_{i}\Pi_{k_{rs}}^{a}+\sum_{i=B_{d}+1}^{B_{d}+d'}c_{i0}X'_{i}$ and thus $\sum_{r=1}^{d}\sum_{i=F_{r}}^{B_{d}}c_{i0}(X'_{i}-\sum_{s=1}^{m_{r}}\Pi_{k_{rs}}^{a}X'_{i}\Pi_{k_{rs}}^{a})+\sum_{i=D}^{m^{2}}c_{i0}X'_{i}=0$. Therefore, we conclude that in the latter relation for $\rho-\Pi^{a}(\rho)$, the term $j=0$ vanishes. Now, we define $K_{rsi}:=tr\Pi_{k_{rs}}^{a}X'_{i}$, then
\begin{align}
\Vert\rho -\Pi^{a}(\rho)\Vert^{2}=\sum_{j=1}^{n^{2}-1}(\sum_{i=D}^{m^{2}}c_{ij}^{2}\qquad\qquad\qquad\qquad\notag \\
+\sum_{r=1}^{d}\sum_{i=F_{r}}^{B_{r}}(c_{ij}^{2}
-\sum_{i'=F_{r}}^{B_{r}}\sum_{s=1}^{m_{r}}K_{rsi}K_{rsi'}c_{ij}c_{i'j}))\quad
\label{91}
\end{align}
The Eq. (\ref{6}) and Eq. (\ref{7}) are changed to
\begin{align}
\sum_{i,i'=B_{r-1}+1}^{B_{r}}K_{rsi}K_{rs'i'}X_{i}X_{i'}=\delta_{ss'}\sum_{i=F_{r}}^{B_{r}}K_{rsi}X_{i},
\label{9}
\end{align}
\begin{align}
\sum_{s=1}^{m_{r}}K_{rsi}=0  &  , &&&&   i=B_{r-1}+2, ..., B_{r}.
\label{10}
\end{align}
respectively. As we demanded, the constraints in the Eq. (\ref{6}) and Eq. (\ref{7}) are divided into independent constraints for each degenerate subspace of $\rho^{a}$.

Also, from the construction of matrices $X'_{i}$, we have $K_{rs F_{r}}=1/\sqrt{m_{r}}$. Using this relation along with Eq. (\ref{10}), we conclude that in Eq. (\ref{91}) the term $F_{r}$ vanishes. Finally, we can write
\begin{align}
\Vert\rho -\Pi^{a}(\rho)\Vert^{2}=\sum_{i=D}^{m^{2}}\sum_{j=1}^{n^{2}-1}c_{ij}^{2}\qquad\qquad\qquad\qquad\notag \\
+\sum_{r=1}^{d}(trC_{r}C_{r}^{t}-trK_{r}C_{r}C_{r}^{t}K_{r}^{t}),\quad
\label{11}
\end{align}
where $C_{r}$ is a $(m_{r}^{2}-1)\times(n^{2}-1)$ dimensional matrix with entries $\lbrace c_{ij}: i=B_{r-1}+2, ..., B_{r}\rbrace$ and $K_{r}$ is a $m_{r}\times(m_{r}^{2}-1)$ dimensional matrix with entries $\lbrace K_{rsi}: s=1, ...,m_{r}; i=B_{r-1}+2, ..., B_{r}\rbrace$. Moreover, notice that Eq. (\ref{9}) and Eq. (\ref{10}), in fact, are some constraints for each matrix $K_{r}$ independent of the others. So we conclude that:
\begin{align*}
\max\Vert\rho -\Pi^{a}(\rho)\Vert^{2}=\sum_{i=D}^{m^{2}}\sum_{j=1}^{n^{2}-1}c_{ij}^{2}\qquad\quad\qquad\qquad\qquad\notag \\
+\sum_{r=1}^{d}\max(trC_{r}C_{r}^{t}-trK_{r}C_{r}C_{r}^{t}K_{r}^{t}).\quad
\end{align*}
Consequently, MIN can be written in the form
\begin{align}
N(\rho)=\sum_{r=1}^{d}N_{r}(\rho)+\sum_{i=D}^{m^{2}}\sum_{j=1}^{n^{2}-1}c_{ij}^{2},\qquad\notag\\
N_{r}(\rho)=trC_{r}C_{r}^{t}-\min_{K'_{r}} trK_{r}C_{r}C_{r}^{t}K_{r}^{t}.\quad
\label{12}
\end{align}
So, the evaluation of MIN for an $m\times n$ dimensional state $\rho$, is degraded to finding the MIN in the $m_{r}\times n$ dimensional subspaces of state $\rho$. This method can reduce the calculations in the evaluation of MIN.

As an important result, we obtain an upper bound for MIN from Eq. (\ref{12}). Following the same arguments in Ref.~\cite{3}, we can write a similar upper bound for each $N_{r}(\rho)$. So we have
\begin{align}
N(\rho)\leq\sum_{r=1}^{d}\sum_{i=1}^{m_{r}^{2}-m_{r}}\lambda_{ri}+\sum_{i=D}^{m^{2}}\sum_{j=1}^{n^{2}-1}c_{ij}^{2}.
\label{13}
\end{align}
where $\lbrace\lambda_{ri}: i=1, ..., m_{r}^{2}-1\rbrace$ are the eigenvalues of the matrix $C_{r}C_{r}^{t}$ listed in decreasing order.

In the next section, we illustrate our previous results for $3\times n$ dimensional states $\rho$, as an example.


\section{$3\times n$ dimensional states}
Consider an arbitrary $3\times n$ dimensional $\rho$. As stated before, when $\rho^{a}$ is non-degenerate, the evaluation of $N(\rho)$ is simple i. e. $N(\rho)=\sum_{i=D}^{m^{2}}\sum_{j=1}^{n^{2}-1}c_{ij}^{2}$. Now, we consider the case where one of eigenvalues of $\rho^{a}$ is doubly degenerate i. e. $\rho^{a}=e_{1}\sum_{s=1}^{2}\vert k_{1s}\rangle\langle k_{1s}\vert+e'_{1}\vert k_{1}\rangle\langle k_{1}\vert$. Thus, in this case, $d=1$, $d'=1$, $m_{1}=2$ and $D=6$. Using Eq. (\ref{12}) we have $N(\rho)=N_{1}(\rho)+\sum_{j=1}^{n^{2}-1}\sum_{i=6}^{9}c_{ij}^{2}$. Computing $N_{1}(\rho)$ is equivalent to computing MIN of a $2\times n$ dimensional state $\rho$ which its marginal $\rho^{a}$ is degenerate. The analytical formula of MIN for an arbitrary $2\times n$ dimensional state $\rho$ has been given in Eq. (\ref{7}) of Ref~\cite{3}. Using this equation we have $N_{1}(\rho)=trC_{1}C_{1}^{t}-\lambda_{min}$ where $\lambda_{min}$ is the smallest eigenvalue of matrix $C_{1}C_{1}^{t}$. In addition, in this case, the upper bound in Eq. (\ref{13}) is equal to $N(\rho)$.

So, when one of eigenvalues of $\rho^{a}$ is doubly degenerate, our method leads to the exact value of the MIN. Moreover, our upper bound in Eq. (\ref{13}) coincide with the exact value of MIN which is, in general, better than the upper bound introduced in~\cite{3}, i. e. Eq. (\ref{5}). Also note that a similar argument is true for an arbitrary $m\times n$ state $\rho$ for which $m_{r}\leq 2$. In the following, we illustrate this, in an example.

Consider the $3\times n$ dimensional state $\rho$ written in the form of Eq. (\ref{2}) as $\rho=\frac{1}{\sqrt{3n}}\frac{I}{\sqrt{3}}\otimes\frac{I}{\sqrt{n}}+\sqrt{\frac{2}{n}}(x-\frac{1}{3})(\sqrt{3}X_{2}-X_{3}+2X_{5}+2X_{7})\otimes\frac{I}{\sqrt{n}}
+\frac{I}{\sqrt{3}}\otimes y_{1}Y_{1}+\sum_{i=1}^{8}t_{i1}X_{i}\otimes Y_{1}$ where $\frac{1}{6}<x<\frac{5}{12}$ and $y_{1}$ and $\lbrace t_{i1}: i=1, ..., 8\rbrace$ are arbitrary real numbers chosen in a such a way that makes $\rho$ to be a valid density operator. Also, in the computational bases, the operators $\lbrace X_{i}: i=1, ..., 8\rbrace$ are in the following form: $X_{1}=\frac{1}{\sqrt{2}}(\vert 1\rangle\langle 1\vert-\vert 2\rangle\langle 2\vert)$, $X_{2}=\frac{1}{\sqrt{6}}(\vert 1\rangle\langle 1\vert+\vert 2\rangle\langle 2\vert-2\vert 3\rangle\langle 3\vert)$, $X_{3}=\frac{1}{\sqrt{2}}(\vert 1\rangle\langle 2\vert+\vert 2\rangle\langle 1\vert)$, $X_{4}=\frac{-i}{\sqrt{2}}(\vert 1\rangle\langle 2\vert-\vert 2\rangle\langle 1\vert)$, $X_{5}=\frac{1}{\sqrt{2}}(\vert 1\rangle\langle 3\vert+\vert 3\rangle\langle 1\vert)$, $X_{6}=\frac{-i}{\sqrt{2}}(\vert 1\rangle\langle 3\vert-\vert 3\rangle\langle 1\vert)$, $X_{7}=\frac{1}{\sqrt{2}}(\vert 2\rangle\langle 3\vert+\vert 3\rangle\langle 2\vert)$ and $X_{8}=\frac{-i}{\sqrt{2}}(\vert 2\rangle\langle 3\vert-\vert 3\rangle\langle 2\vert)$. Using Eq. (\ref{5}) we have $N(\rho)\leq\sum_{i=1}^{8}t_{i1}^{2}$. We now evaluate our upper bound as follows. At first, we evaluate the reduced density matrix of $\rho$. The spectral decomposition of $\rho^{a}$ is $\rho^{a}=e_{1}\sum_{s=1}^{2}\vert k_{1s}\rangle\langle k_{1s}\vert+e'_{1}\vert k_{1}\rangle\langle k_{1}\vert$ where $e_{1}=2x-1/3$, $e'_{1}=-4x+5/3$, $\vert k_{11}\rangle=\frac{1}{\sqrt{3}}(\vert 1\rangle+\vert 2\rangle+\vert 3\rangle)$,  $\vert k_{12}\rangle=\frac{1}{\sqrt{2}}(\vert 1\rangle-\vert 2\rangle)$ and $\vert k_{1}\rangle=\frac{1}{\sqrt{6}}(\vert 1\rangle+\vert 2\rangle-2\vert 3\rangle)$. We now construct the $\lbrace X'_{i}: i=1, ...,9\rbrace$ which are used in the expansion of $\rho$ in the form of Eq. (\ref{8}). We construct $\lbrace X'_{i}: i=1, ...,4\rbrace$ as follows; $X'_{1}=\frac{1}{\sqrt{2}}(\vert k_{11}\rangle\langle k_{11}\vert+\vert k_{12}\rangle\langle k_{12}\vert)$, $X'_{2}=\frac{1}{\sqrt{2}}(\vert k_{11}\rangle\langle k_{12}\vert+\vert k_{12}\rangle\langle k_{11}\vert)$, $X'_{3}=\frac{-i}{\sqrt{2}}(\vert k_{11}\rangle\langle k_{12}\vert-\vert k_{12}\rangle\langle k_{11}\vert)$ and $X'_{4}=\frac{1}{\sqrt{2}}(\vert k_{11}\rangle\langle k_{11}\vert-\vert k_{12}\rangle\langle k_{12}\vert)$. In addition we set $X'_{5}=\vert k_{1}\rangle\langle k_{1}\vert$. The operators $\lbrace X'_{i}: i=6, ...,9\rbrace$ are constructed in the following form; $X'_{6}=\frac{1}{\sqrt{2}}(\vert k_{11}\rangle\langle k_{1}\vert+\vert k_{1}\rangle\langle k_{11}\vert)$, $X'_{7}=\frac{-i}{\sqrt{2}}(\vert k_{11}\rangle\langle k_{1}\vert-\vert k_{1}\rangle\langle k_{11}\vert)$, $X'_{8}=\frac{1}{\sqrt{2}}(\vert k_{12}\rangle\langle k_{1}\vert+\vert k_{1}\rangle\langle k_{12}\vert)$, $X'_{9}=\frac{-i}{\sqrt{2}}(\vert k_{12}\rangle\langle k_{1}\vert-\vert k_{1}\rangle\langle k_{12}\vert)$.

Now, we write $\rho$ in the form of Eq. (\ref{8}) using $\lbrace X'_{i}: i=1, ...,9\rbrace$. We have $\rho=(\frac{\sqrt{2}e_{1}}{\sqrt{n}}X'_{1}+\frac{e'_{1}}{\sqrt{n}}X'_{5})\otimes\frac{1}{\sqrt{n}}
+\sum_{i=1}^{9}c_{i1}X'_{i}\otimes Y_{1}$ where $c_{11}=\frac{2}{\sqrt{6}}t_{11}+\frac{1}{\sqrt{12}}t_{31}-\frac{1}{6}t_{41}+\frac{1}{3}(t_{61}+t_{81})$, $c_{21}=\frac{1}{\sqrt{6}}(2t_{21}+t_{61}-t_{81})$, $c_{31}=\frac{1}{\sqrt{6}}(-2t_{51}-t_{71}+t_{91})$, $c_{41}=-\frac{1}{\sqrt{12}}t_{31}+\frac{5}{6}t_{41}+\frac{1}{3}(t_{61}+t_{81})$, $c_{51}=\frac{1}{\sqrt{3}}t_{11}-\frac{1}{\sqrt{6}}t_{31}+\frac{\sqrt{2}}{6}t_{41}-\frac{\sqrt{2}}{3}(t_{61}+t_{81})$, $c_{61}=\frac{2}{\sqrt{6}}t_{31}+\frac{\sqrt{2}}{3}t_{41}-\frac{\sqrt{2}}{6}(t_{61}+t_{81})$, $c_{71}=\frac{-1}{\sqrt{2}}(t_{71}+t_{91})$, $c_{81}=\frac{1}{\sqrt{3}}(t_{21}-t_{61}+t_{81})$ and $c_{91}=\frac{1}{\sqrt{3}}(t_{51}-t_{71}+t_{91})$. According to our previous descriptions (in the first paragraph of this section), we obtain $N_{1}(\rho)=\sum_{i=2}^{4}c_{i1}^{2}$. So $N(\rho)=-c_{51}^{2}+\sum_{i=2}^{9}c_{i1}^{2}$. As stated before, in this case our upper bound is equal to $N(\rho)$. The difference between the upper bound introduced in Eq. (\ref{5}) and our upper bound is equal to $\frac{1}{6}(\frac{3}{\sqrt{6}}t_{31}+\sqrt{2}t_{61}+\sqrt{2}t_{81}-\frac{1}{\sqrt{2}}t_{41})^{2}$. So our upper bound is better than the upper bound introduced in Eq. (\ref{5}).

Now consider the case that $\rho^{a}$ is fully degenerate i. e. $\rho^{a}$ have three identical eigenvalues. In this case, we have $d=1$, $d'=0$, $m_{1}=m=3$ and so Eqs. (\ref{9}), (\ref{10}), (\ref{12}) are the same as Eqs. (\ref{6}), (\ref{7}), (\ref{3}), respectively. We use the notation of Eqs. (\ref{6}), (\ref{7},\ref{3}) for simplicity. When $\rho^{a}$ is fully degenerate, then $\rho^{a}=\frac{I}{m}$. So the constraint $\sum_{k}\Pi_{k}^{a}\rho^{a}\Pi_{k}^{a}=\rho^{a}$ set no conditions on $\lbrace\Pi_{k}^{a}\rbrace$. Therefore, in the optimization problem of Eq. (\ref{3}), only the constraints in the Eqs. (\ref{6}), (\ref{7}) are present. Using the $\lbrace X_{i}\rbrace$ represented in the previous paragraph, we can obtain the constraints on $\lbrace a_{ki}\rbrace$ which can be derived from Eqs. (\ref{6}), (\ref{7}). These constraints are represented in the appendix.

In order to evaluate MIN, we must find the minimum of $trATT^{t}A^{t}$ with the constraints obtained from the Eq. (\ref{6}) and Eq. (\ref{7}) in the previous paragraphs. But, if we put additional constraints on the $\lbrace a_{ki}\rbrace$ simplifying the previous represented constraints, then we can find some analytical lower bounds for MIN. For example if we put $a_{11}=a_{21}=0$, $a_{12}=a_{22}=0$ and $a_{26}=a_{28}=0$ then we obtain $a_{31}=a_{32}=0$, $a_{25}=a_{27}=a_{13}=a_{23}=\frac{\sqrt{2}}{3}$, $a_{33}=\frac{-2\sqrt{2}}{3}$, $a_{15}=a_{17}=a_{35}=a_{37}=\frac{-1}{3\sqrt{2}}$, $a_{16}=a_{18}=\frac{1}{\sqrt{6}}$,  $a_{36}=a_{38}=\frac{-1}{\sqrt{6}}$ and $a_{14}=a_{24}=a_{34}=0$ and thereby we obtain a lower bound for MIN. For every state $\rho$ with $\lbrace t_{1j}=t_{2j}=t_{4j}=0: j=1, ..., n^{2}-1\rbrace$, this lower bound is equal to the exact value of MIN. So depending on the coefficients $t_{ij}$ in the expansion of the the state $\rho$, this lower bound can be a good lower bound for MIN. According to this method, we can find good lower bounds for MIN of the other classes of states.


\section{conclusions}
We have studied the evaluation of MIN for an arbitrary $m\times n$ dimensional bipartite density matrix $\rho$. We have shown that according to the degeneracy of $\rho^{a}$, expanding $\rho$ in the suitable basis, the evaluation of MIN for an $m\times n$ dimensional state $\rho$, is degraded to finding the MIN in the $m_{i}\times n$ ($  m_{i} \leq m $) dimensional subspaces of state $\rho$. This method can reduce the calculations in the evaluation of MIN. Moreover, for an arbitrary $m\times n$ state $\rho$ for which $m_{i}\leq 2$, our method leads to the exact value of the MIN. Also, we obtain an upper bound for MIN which can improve the ones introduced in Ref.~\cite{3}. In addition, we have explained the evaluation of MIN for $3\times n$ dimensional states and introduced some lower bounds for MIN when $\rho^{a}$ is fully degenerate.


\section*{acknowledgments}
One of authors, S. Y. M, would like to thank Emad Ebrahimi, Fateme Ghomanjani and Roholamin Zeinali for useful discussions. Also, S. Y. M acknowledge financial support from the Vali-e-Asr University of Rafsanjan.


\appendix*
\renewcommand{\theequation}{A.\arabic{equation}}
\setcounter{equation}{0}
\section*{APPENDIX}
In this appendix, we represent the constraints on $\lbrace a_{ki}\rbrace$ which can be derived from Eqs. (\ref{6}), (\ref{7}).
From Eq. (\ref{6}), for $k'=k=1, 2$, we obtain (we recall that $a_{10}=a_{20}=\frac{1}{\sqrt{3}}$)
\begin{flushleft}
$\sum_{i=1}^{8}a_{ki}^{2}=\frac{2}{3}$\\
$\frac{1}{3}a_{k1}=\frac{2}{\sqrt{6}}a_{k1}a_{k2}+\frac{1}{2\sqrt{2}}(a_{k5}^{2}+a_{k6}^{2}-a_{k7}^{2}-a_{k8}^{2}),$\\
$\frac{1}{3}a_{k2}=\frac{1}{\sqrt{6}}(a_{k1}^{2}-a_{k2}^{2}+a_{k3}^{2}+a_{k4}^{2}-\frac{1}{2}(a_{k5}^{2}+a_{k6}^{2}+a_{k7}^{2}+a_{k8}^{2}))$,\\
$\frac{1}{3}a_{k3}=\frac{2}{\sqrt{6}}a_{k3}a_{k2}+\frac{1}{\sqrt{2}}(a_{k5}a_{k7}+a_{k6}a_{k8})$,\\
$\frac{1}{3}a_{k4}=\frac{2}{\sqrt{6}}a_{k4}a_{k2}+\frac{1}{\sqrt{2}}(-a_{k5}a_{k8}+a_{k6}a_{k7})$,\\
$\frac{1}{3}a_{k5}=\frac{-1}{\sqrt{6}}a_{k5}a_{k2}+\frac{1}{\sqrt{2}}(a_{k1}a_{k5}+a_{k3}a_{k7}-a_{k4}a_{k8})$,\\
$\frac{1}{3}a_{k6}=\frac{-1}{\sqrt{6}}a_{k6}a_{k2}+\frac{1}{\sqrt{2}}(a_{k1}a_{k6}+a_{k3}a_{k8}+a_{k4}a_{k7})$,\\
$\frac{1}{3}a_{k7}=\frac{-1}{\sqrt{6}}a_{k7}a_{k2}+\frac{1}{\sqrt{2}}(-a_{k1}a_{k7}+a_{k3}a_{k5}+a_{k4}a_{k6})$,\\
$\frac{1}{3}a_{k8}=\frac{-1}{\sqrt{6}}a_{k8}a_{k2}+\frac{1}{\sqrt{2}}(-a_{k1}a_{k8}+a_{k3}a_{k6}-a_{k4}a_{k5})$.
\end{flushleft}
In addition, if we put $k=1$ and $k'=2$  in the Eq. (\ref{6}), then the following constraints on $\lbrace a_{ki}\rbrace$ can be obtained.
$\sum_{i=1}^{8}a_{1i}a_{2i}=\frac{-1}{3}$,\newline

$\frac{1}{3}(a_{11}+a_{21})+\frac{1}{\sqrt{6}}(a_{11}a_{22}+a_{12}a_{21})+\frac{1}{2\sqrt{2}}(a_{15}a_{25}+a_{16}a_{26}-a_{17}a_{27}-a_{18}a_{28})+\frac{i}{2\sqrt{2}}(2a_{14}a_{23}-2a_{13}a_{24}+a_{16}a_{25}-a_{15}a_{26}+a_{17}a_{28}-a_{18}a_{27})=0$,\newline

$\frac{1}{3}(a_{12}+a_{22})+\frac{1}{\sqrt{6}}(a_{11}a_{21}-a_{12}a_{22}+a_{13}a_{23}+a_{14}a_{24}-\frac{1}{2}a_{15}a_{25}-\frac{1}{2}a_{16}a_{26}-\frac{1}{2}a_{17}a_{27}-\frac{1}{2}a_{18}a_{28})+\frac{3i}{2\sqrt{6}}(a_{16}a_{25}-a_{15}a_{26}+a_{18}a_{27}-a_{17}a_{28})=0$,\newline

$\frac{1}{3}(a_{13}+a_{23})+\frac{1}{\sqrt{6}}(a_{12}a_{23}+a_{13}a_{22})+\frac{1}{2\sqrt{2}}(a_{15}a_{27}+a_{17}a_{25}+a_{16}a_{28}+a_{18}a_{26})+\frac{i}{\sqrt{2}}(a_{11}a_{24}-a_{14}a_{21}-\frac{1}{2}a_{15}a_{28}+\frac{1}{2}a_{18}a_{25}+\frac{1}{2}a_{16}a_{27}-\frac{1}{2}a_{17}a_{26})=0$,\newline

$\frac{1}{3}(a_{14}+a_{24})+\frac{1}{\sqrt{6}}(a_{12}a_{24}+a_{14}a_{22})+\frac{1}{2\sqrt{2}}(-a_{15}a_{28}-a_{18}a_{25}+a_{16}a_{27}+a_{17}a_{26})+\frac{i}{\sqrt{2}}(a_{13}a_{21}-a_{11}a_{23}-\frac{1}{2}a_{15}a_{27}+\frac{1}{2}a_{17}a_{25}+\frac{1}{2}a_{18}a_{26}-\frac{1}{2}a_{16}a_{28})=0$,\newline

$\frac{1}{3}(a_{15}+a_{25})-\frac{1}{2\sqrt{6}}(a_{12}a_{25}+a_{15}a_{22})+\frac{1}{2\sqrt{2}}(a_{11}a_{25}+a_{15}a_{21}+a_{13}a_{27}+a_{17}a_{23}-a_{14}a_{28}-a_{18}a_{24})+\frac{i}{2\sqrt{2}}(a_{11}a_{26}-a_{16}a_{21}+a_{13}a_{28}-a_{18}a_{23}+a_{14}a_{27}-a_{17}a_{24}+\sqrt{3}a_{12}a_{26}-\sqrt{3}a_{16}a_{22})=0$,\newline

$\frac{1}{3}(a_{16}+a_{26})-\frac{1}{2\sqrt{6}}(a_{12}a_{26}+a_{16}a_{22})+\frac{1}{2\sqrt{2}}(a_{11}a_{26}+a_{16}a_{21}+a_{13}a_{28}+a_{18}a_{23}+a_{14}a_{27}+a_{17}a_{24})+\frac{i}{2\sqrt{2}}(-a_{11}a_{25}+a_{15}a_{21}-a_{13}a_{27}+a_{17}a_{23}+a_{14}a_{28}-a_{18}a_{24}+\sqrt{3}a_{15}a_{22}-\sqrt{3}a_{12}a_{25})=0$,\newline

$\frac{1}{3}(a_{17}+a_{27})-\frac{1}{2\sqrt{6}}(a_{12}a_{27}+a_{17}a_{22})+\frac{1}{2\sqrt{2}}(-a_{11}a_{27}-a_{17}a_{21}+a_{13}a_{25}+a_{15}a_{23}+a_{14}a_{26}+a_{16}a_{24})+\frac{i}{2\sqrt{2}}(-a_{11}a_{28}+a_{18}a_{21}+a_{13}a_{26}-a_{16}a_{23}-a_{14}a_{25}+a_{15}a_{24}+\sqrt{3}a_{12}a_{28}-\sqrt{3}a_{18}a_{22})=0$ and \newline

$\frac{1}{3}(a_{18}+a_{28})-\frac{1}{2\sqrt{6}}(a_{12}a_{28}+a_{18}a_{22})+\frac{1}{2\sqrt{2}}(-a_{11}a_{28}-a_{18}a_{21}+a_{13}a_{26}+a_{16}a_{23}-a_{14}a_{25}-a_{15}a_{24})+\frac{i}{2\sqrt{2}}(a_{11}a_{27}-a_{17}a_{21}-a_{13}a_{25}+a_{15}a_{23}-a_{14}a_{26}+a_{16}a_{24}+\sqrt{3}a_{17}a_{22}-\sqrt{3}a_{12}a_{27})=0$.

From Eq. (\ref{7}), for $i=1, ..., m^{2}$, we obtain $a_{3i}=-(a_{1i}+a_{2i})$. Using this relation, it is concluded that by choosing $k, k'=3$ in the Eq. (\ref{6}), no additional constraint can be obtained.



\begin{thebibliography}{99}
\bibitem{5} We note that for the construction of $\Pi^{a}$, we are not restricted to choose the $\vert k'_{rs}\rangle$ used in the construction of $X'_{i}$.
\end{thebibliography}
\end{document}